\documentclass[prd,preprint]{revtex4}
\usepackage{graphicx}
\begin{document}

\newcommand{\re}{\mathop{\mathrm{Re}}}

\newcommand{\be}{\begin{equation}}
\newcommand{\ee}{\end{equation}}
\newcommand{\bea}{\begin{eqnarray}}
\newcommand{\eea}{\end{eqnarray}}

%\maketitle
 
\title{Brane universes tested by supernovae Ia}

\author{W{\l }odzimierz God{\l }owski}
\email{godlows@oa.uj.edu.pl}
\affiliation{\it Astronomical Observatory, Jagiellonian University, 30-244
Krakow, ul. Orla 171, Poland}
\author{Marek Szyd{\l }owski}
\email{uoszydlo@cyf-kr.edu.pl}
\affiliation{\it Astronomical Observatory, Jagiellonian University, 30-244
Krakow, ul. Orla 171, Poland}
 
\begin{abstract}
We discuss observational constrains coming from supernovae Ia
imposed on the behaviour of the Randall-Sundrum models. In the case
of dust matter on the brane, the difference between the best-fit Perlmutter
model with a $\Lambda$-term and the best-fit
brane models becomes detectable for redshifts $z > 1.2$. It is interesting that
brane models predict brighter galaxies for such redshifts which
is in agreement with the measurement of the $z = 1.7$ supernova.
We also demonstrate that the fit to supernovae data can also be
obtained, if we admit the "super-negative" dark energy (phantom matter)
$p = - (4/3) \varrho$ on the brane,
where the dark energy in a way mimics the influence of the cosmological
constant. It also appears that the dark energy enlarges the age of the universe
which is demanded in cosmology. Finally, we propose to check for dark radiation
and brane tension by the application of the angular diameter of galaxies minimum
value test. We point out the existence of coincidence problem for the brane
tension parameter.
\end{abstract}
 
\maketitle
 
\section{Introduction}
 
In recent several years a lot of effort has been done on the idea that
our Universe is a boundary of a higher-dimensional spase time manifold
\cite{Arkani-Hamed98,Arkani-Hamed99}. Kaluza and Klein first discussed
$5-dimensional$ space time to unify a gravity and electromagnetism.
Among supersting theories which may unify all interactions M-theory
is a strong candidate for the description of real world. In this theory,
gravity is a truly higher-dimensional theory, becoming effectively
4-dimensional at lower energies. The standard model matter fields are confined
to the 3-brane while gravity can, by its universal character, propagate
in all extra dimensions. In the brane world models inspired by string/M theory
(\cite{rs1},\cite{rs2}, \cite{hw}) new two parameters which
doesn't present in standard cosmology are introduced,
namely brane tension $\lambda$ and dark radiation $U$. One of new
approaches was proposed by Randall and Sundrum \cite{rs1}, \cite{rs2}
where our Minkowski brane is localized in 5dimensional anti-de Sitter space
time with metric:
\begin{equation}
\label{eq:1}
ds^2=\exp(-2|y|l)(-dt^2+d\vec{x}^2)+dy^2.
\end{equation}

For $y \ne 0$, this metric satisfies the 5dimensional Einstein equation with the
negative cosmological constant $\tilde{\check{\Lambda_{(5)}}} \propto -l^{-2}$.
The brane is located at $y=0$ and the induced metric on brane is a Minkowski
metric. The bulk is a 5-dimensional anti-deSitter metric with $y=0$ as a boundary.
 
We should mention that before the Randall and Sundrum work \cite{rs2} where
they proposed a mechanism to solve the hierarchy problem by a small extra
dimension, large extra dimensions were proposed to solve this problem by
Arkani-Hamed et.al. \cite{Arkani-Hamed98,Arkani-Hamed99}.
This gives an interesting feature because TeV gravity might be realistic and
quantum gravity effects could be observed by a next generation particle collider.
The Newtonian gravity potential on the brane is recovered at lowest order
$V(r)=\frac{GM}{r}(1+\frac{2l^2}{3r^2})$. In this paper we demonstrate that
if the brane world is the Randall-Sundrum version is realistic we may find some
evidence of higher dimensions.

In  \cite{Szydlo02} we gave the formalism to express dynamical equations
in terms of dimensionless observational density parameters $\Omega$.
In this notation (see also \cite{Dabrowski96,AJI+II,AJIII})
the Friedmann equation for brane universes takes the form
\begin{equation}
\label{eq:FriedCCC}
\frac{1}{a^2} \left( \frac{da}{dt} \right)^2 =
\frac{C_{\gamma}}{a^{3\gamma}} + \frac{C_{\lambda}}{a^{6\gamma}} -
\frac{k}{a^2} + \frac{\Lambda_{(4)}}{3} + \frac{C_{\cal U}}{a^4} ,
\end{equation}
where $a(t)$ is the scale factor, $k=0,\pm1$ the curvature index,
here we use natural system of units in which $8\pi G=c=1$, $\Lambda_{(4)}$ is
the 4-dimensional cosmological contant, and $\gamma$ the barotropic index
($p = (\gamma - 1)\varrho$, $p$ - the pressure, $\varrho$ - the energy
density), the constants
$C_{\lambda} = 1/6\lambda \cdot a^{6\gamma}
\varrho^2$ and $C_{\cal U} = 2/ \lambda \cdot a^4 {\cal U}$,
$C_{\lambda}$ comes as a contribution from brane tension
$\lambda$, and $C_{\cal U}$ as a contribution from dark radiation.
 
Because $\rho^2$ term and dark radiation term do not appear in the
standard cosmology, such terms could provide a smal window to see the extra
dimensions.
 
In order to study observational tests we now define
dimensionless observational density parameters
\bea
\label{Omegadef}
\Omega_{\gamma}  &=&  \frac{1}{3H^2} \varrho ,
\hspace{15pt}
\Omega_{\lambda}  =  \frac{1}{6H^2\lambda} \varrho^2 ,
\hspace{15pt}
\Omega_{\cal U}  =  \frac{2}{H^2\lambda} {\cal U}
,\nonumber \\
\Omega_{k}  &=&  - \frac{k}{H^2a^2} ,
\hspace{15pt}
\Omega_{\Lambda_{(4)}}  =  \frac{\Lambda_{(4)}}{3H^2} ,
\eea
where the Hubble parameter $H = \dot{a}/a$, and the deceleration parameter
$q  =  - \ddot{a}a/\dot{a}^2$ ,
so that the Friedmann equation (\ref{eq:FriedCCC}) can be written down
in the form
\begin{equation}
\label{Om=1}
\Omega_{\gamma} + \Omega_{\lambda} + \Omega_{k} + \Omega_{\Lambda_{(4)}} + \Omega_{\cal U}
= 1  .
\end{equation}
Note that $\Omega_{\cal U}$ in (\ref{Omegadef}), despite standard radiation term,
can either be positive or negative.
 
It is useful to rewrite (\ref{eq:FriedCCC}) to the dimensionless form.
Let us consider a standard Friedmann-Robertson-Walker universe (hereafter FRW)
filled with mixture of matter with the equation of state
$p_i = (\gamma_i-1) \rho_i$. Then we obtain the basic equation in the form:
 
\begin{equation}
\label{eq:10}
\frac{\dot{x}^{2}}{2} = \frac{1}{2} \Omega_{k,0} + \frac{1}{2}\sum_{i}
\Omega_{i,0} x^{2-3\gamma_{i}} =-V(x)\\
\end{equation}
\begin{equation}
\label{eq:10a}
\ddot{x} = - \frac{1}{2} \sum_{i} \Omega_{i,0}(2-3\gamma_{i})
x^{1-3\gamma_{i}}=-\frac{\partial V(x)}{\partial x}
\end{equation}
where $i=(\gamma,\lambda,\Lambda,U)$,
and
 
\begin{equation}
\label{eq:11}
x \equiv \frac{a}{a_{0}}, \qquad T \equiv |H_{0}| t,
\qquad \dot{}\equiv \frac{d}{dT}.
\end{equation}
and $t$ is original cosmological time, $V$ is the potential function.
 
Therefore the dynamics of the considered model is equivalent
to introducing fictitious fluids which mimic $\rho^2$ contribution and dark
energy term. For dark energy $\gamma=4/3$ whereas for brane $\gamma_{\lambda}=
2\gamma$. The presented formalism is useful in analysis of observational
tests of brane models.

Above relations allow to write down an explicit redshift-magnitude
relation  for the brane models to study
their compatibility with astronomical data which
is the subject of the present paper. Obviously, the luminosity of galaxies
depends on the present densities of the different components
of matter content $\Omega_{i}$ given by (\ref{Omegadef}) and their equations of
state, reflected by the value of the barotropic index $\gamma_i$.
 
\section{Brane cosmologies and SNIa observations}
 
Let us consider an observer located at $r=0$ at the moment $t=t_0$
which receives a light ray emitted at $t=t_1$ from the source of the absolute luminosity $L$
located at the radial distance $r_1$. The redshift $z$
of the source is related to the scale factor $a(t)$ at the two moments of evolution
by $1+z=a(t_0)/a(t_1) \equiv a_0/a$.
If the apparent luminosity of the source as measured by the observer is $l$,
then the luminosity distance $d_L$ of the source is defined by the
relation
\be
\label{luminosity}
l={L\over 4\pi d_L^2},
\ee
where
\be
\label{DeeL}
d_L=(1+z)a_0 r_1 \equiv \frac{{\cal D}_L(z)}{H_0},
\ee
and ${\cal D}_L$ is the dimensionless luminosity distance.
The observed and absolute luminosities are
defined in terms of K-corrected apparent and absolute
magnitudes $m$ and $M$. When written in terms of $m$ and $M$,
Eq. (\ref{luminosity}) yields
\be
\label{m(z)}
m(z)={\cal M} + 5\log_{10}
[{\cal D}_L(z)],
\ee
where ${\cal M}=M-5\log_{10}H_0+25$.
For homogeneous and isotropic Friedmann models one gets
\be
\label{Deelfin}
{\cal D}_L(z) = \frac{\left( 1+z \right)}
{\sqrt{{\cal K}}} S(\chi)
\ee
where
$S(\chi)=\sin \chi$ with ${\cal K}=-\Omega_{k,0}$ when $\Omega_{k,0}<0$;
$S(\chi)=\chi $ with ${\cal K}=1$ when $\Omega_{k,0}=0$;
$S(\chi)=\sinh \chi$ with ${\cal K}=\Omega_{k,0}$  when $\Omega_{k,0}>0$.
From the Friedmann equation (\ref{eq:FriedCCC}) and the form of the FRW metric
we have
\bea
\label{chir1}
\chi(z)=\sqrt{{\cal K}}\int\limits_0^z
\left\{\Omega_{\lambda,0}\left(1+z^{'} \right)^{6\gamma} +
\Omega_{\gamma,0}\left(1+z^{'} \right)^{3\gamma}
\right. \nonumber \\ \left. + \Omega_{k,0}\left(1+z^{'} \right)^2 +
\Omega_{{\cal U},0}\left(1+z^{'}\right)^4 +
\Omega_{\Lambda_{(4),0}}\right\}^{-1/2}dz^{'}.
\eea
Firstly, we will study the case $\gamma = 1$ (dust on the brane;
we will label $\Omega_{\gamma}$ by $\Omega_m $).
The case $\gamma = 2/3$
(cosmic strings on the brane) has recently been studied in
\cite{Singh1/3} where, in fact,
$\Omega_{\cal U}$ and $\Omega_{\lambda}$ were neglected and
where the term $\Omega_{m,0} (1+z^{'})^3$ was
introduced in order to admit dust matter on the brane.
This case was already presented in
a different framework in Ref. \cite{AJIII}.
Secondly, we will study the case $\gamma = -1/3$
(dark energy on the brane \cite{darkenergy}
- we will label this type of matter with $\Omega_{d}$ instead of
$\Omega_{\gamma}$).
 
Now we test brane models using the Perlmutter samples \cite{Perlmutter99}.
In order to avoid any possible selection effects, we use the full sample
(usually, one excludes
two data points as outliers and another two points, presumably reddened,
from the full sample of 60 supernovae). It means that our basic sample is
the Perlmutter sample A \cite{Perlmutter99}.
We test our model using the likelihood method \cite{Riess98}.

First of all, we estimate the value of $\mathcal{M}$
from the full sample of 60 supernovae. For the flat model we obtained
$\mathcal{M} = -3.39$ which is in agreement with the
results of \cite{Efstathiou99,Vishwakarma}.
Also, we obtained for the Perlmutter model the same value of
$\chi^2 = 96.5$ what is in agreement with \cite{Perlmutter99}.
 
Neglecting dark radiation $\Omega_{{\cal U},0} = 0$ we formally got
the best fit  ($\chi^2=94.6$) for $\Omega_{k,0}=-0.9$,
$\Omega_{m,0}=0.59$, $\Omega_{\lambda,0}=0.04$, $\Omega_{\Lambda,0}=1.27$,
(see Tab I) which is completely unrealistic \cite{Peebles02,Lahav02},
because $\Omega_{m,0}=0.59$ is too large in
comparison with the observational limit (also $\Omega_{k,0}=-0.9$
is not very realistic from the observational point of view).
 
\begin{figure}[h]
\includegraphics[width=0.8\textwidth]{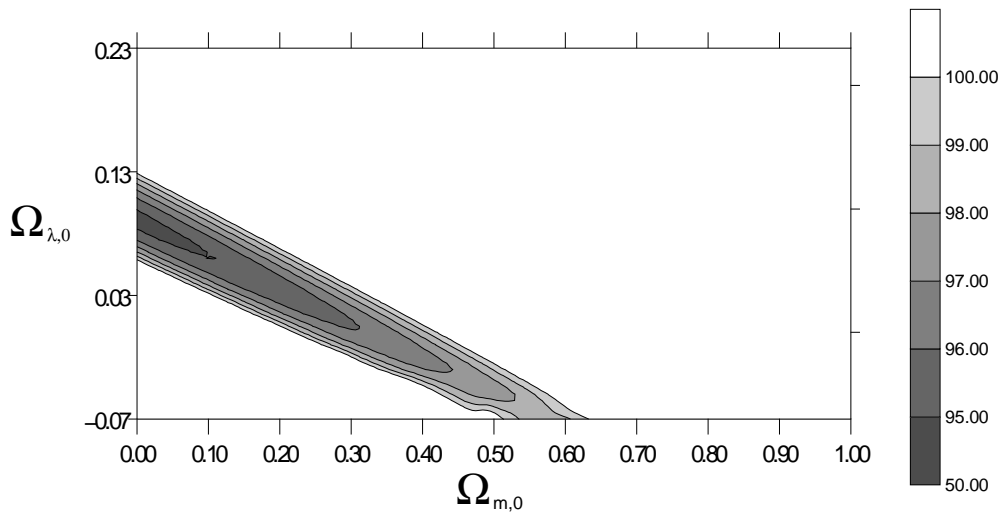}
\caption{The plot of $\chi^2$ levels for flat ($\Omega_{k,0} = 0$) brane models with
respect to the values of $\Omega_{m,0}$ (horizontal axis) and $\Omega_{\lambda,0}$
(vertical axis).}
\label{Fig.1}
\end{figure}

However, we should note that, in fact, we have an ellipsoid of admissible
models in a 3dimensional parameter space $\Omega_{m,0}$, $\Omega_{\lambda,0}$,
$\Omega_{\Lambda_{(4)},0}$ at hand. Then, we have more freedom than in the
case of analysis of \cite{Perlmutter99} where they had only an ellipse in a
2dimensional parameter space $\Omega_{m,0}$, $\Omega_{\Lambda_{(4)},0}$.
For a flat model $\Omega_{k,0}=0$
we obtain "corridors" of possible models (Fig.\ref{Fig.1}).
Formally, the best-fit flat model is $\Omega_{m,0}=0.01$,
$\Omega_{\lambda,0}=0.09$ $\chi^2=94.7$ which is again unrealistic.
In the realistic case we obtain for a flat model $\Omega_{m,0}=0.25$,
$\Omega_{\lambda,0}=0.02$, $\Omega_{\Lambda_{(4)},0}=0.73$
with $\chi^2=95.6$. One should note that all
realistic brane models also require the presence of the positive
4-dimensional cosmological constant ($\Omega_{\Lambda_{(4)},0} \sim 0.7$).

There is a question if we could fit a model with negative
$\Omega_{\lambda,0}$? For instance, in a flat Universe we could
fit the model with $\Omega_{m,0}=0.35$ (too much in comparison
with the observational limits on the mass of the cluster of galaxies)
$\Omega_{\lambda,0}=-0.01$, $\Omega_{\Lambda,0}=0.66$ ($\chi^2=96.3$).

\begin{figure}
\includegraphics[width=0.8\textwidth]{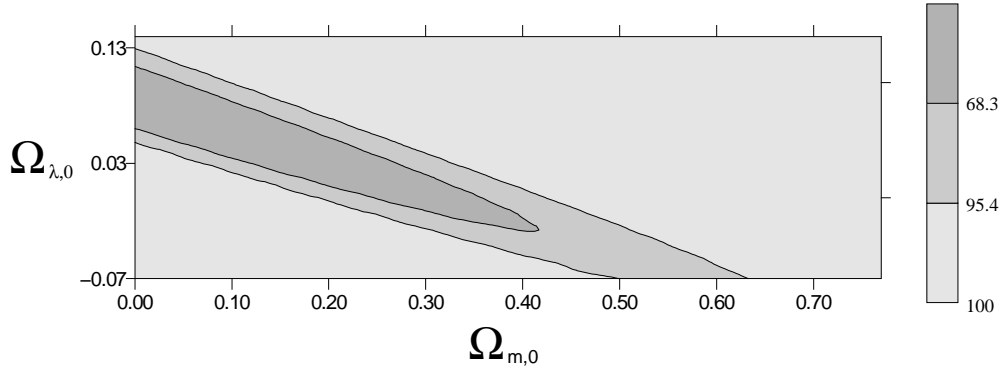}
\caption{Confidence levels on the plane
$(\Omega_{\mathrm{m},0} , \Omega_{\lambda,0})$ minimalized over $\mathcal{M}$
for the flat model, and with
$\Omega_{\Lambda,0}=1-\Omega_{\mathrm{m},0}-\Omega_{k,0}-\Omega_{\lambda,0}$.
The figure shows the ellipses of the preferred value of
$\Omega_{\mathrm{m},0}$ and $\Omega_{\Lambda,0}$.}
\label{Fig:2}
\end{figure}

Fig.~\ref{Fig:2} illustrates the confidence level as a function of
($\Omega_{\mathrm{m},0}$, $\Omega_{\lambda,0}$) for the flat model
($\Omega_{k,0}=0$) minimalized over $\mathcal{M}$ with
$\Omega_{\Lambda,0}=1-\Omega_{\mathrm{m},0}-\Omega_{k,0}-\Omega_{\lambda,0}$.
In present cases we formally assumed that both positive and negative value of
$\Omega_{\lambda,0}$ are mathematically possible.
We show that the preferred intervals for $\Omega_{m,0}$ and
$\Omega_{\lambda,0}$ are $\Omega_{m,0}<0.4$ and
$\Omega_{\lambda,0}>0$.
 
\begin{figure}[h]
\includegraphics[width=0.8\textwidth]{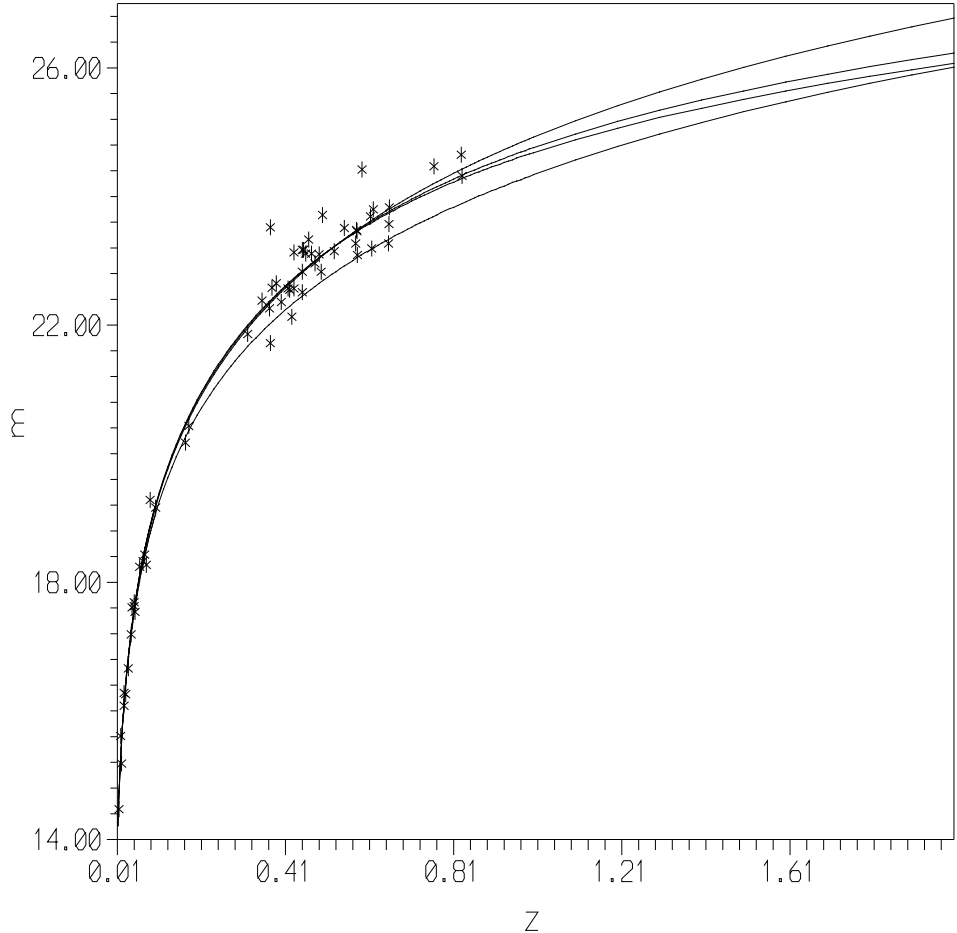}
%\includegraphics[scale=.31]{ff3.eps}
%\centerline{\epsfxsize=4cm\epsfbox{moddab.ps}}
\caption{The redshift-magnitude relation for $\gamma = 1$ brane universes
(dust on the brane). The top line is the best-fit Perlmutter model.
with $\Omega_{m,0}=0.28$, $\Omega_{\Lambda_{(4)},0}=0.72$.
The bottom line is a flat model with $\Omega_{m,0}=1$. Between these
two lines there are brane models with $\Omega_{\lambda,0} \ne 0$: lower--the
best-fit nonflat model; higher--the best-fit flat model.}
\label{Fig.3}
\end{figure}
 
\begin{table}
\noindent
\caption{Results of the statistical analysis for the dust matter on the brane
for Perlmutter Sample A, B, C. Two upper line for each sample
are best fit model and best fit flat model for sample. Third line is
"realistic" model with $\Omega_{\text{m},0}\simeq 0.3$. We also include, for
sample A the model with $\Omega_{\lambda,0}<0$
}
\begin{tabular}{@{}p{1.5cm}p{0.5cm}rrrrr}
\hline
Sample & N & $\Omega_{k,0}$ & $\Omega_{\lambda,0}$ & $\Omega_{m,0}$ & $\Omega_{\Lambda,0}$ & $\chi^2$ \\
\hline
  A   &  60 & -0.9 &  0.04 & 0.59 & 1.27 & 94.7\\
      &     &  0.0 &  0.09 & 0.01 & 0.90 & 94.7\\
      &     &  0.0 &  0.02 & 0.25 & 0.73 & 95.6\\
      &     &  0.0 & -0.01 & 0.35 & 0.66 & 96.3\\
      &     &      &       &      &     &     \\
\hline
  B   &  56 & -0.1 &  0.06 & 0.17 & 0.87 & 57.3\\
      &     &  0.0 &  0.06 & 0.12 & 0.82 & 57.3\\
      &     &  0.0 &  0.02 & 0.25 & 0.73 & 57.6\\
      &     &      &       &      &     &     \\
\hline
  C   &  54 &  0.0 &  0.04 & 0.21 & 0.73 & 53.5\\
      &     &  0.0 &  0.04 & 0.21 & 0.73 & 53.5\\
      &     &  0.0 &  0.02 & 0.27 & 0.71 & 53.6
\end{tabular}
\end{table}

In Fig. \ref{Fig.3} we present plots of redshift-magnitude relations against
the supernovae data.
One can observe that in both cases (best-fit and best-fit flat models) the
difference between brane models and a pure flat (Einstein-de Sitter)
model with $\Omega_{\Lambda_{(4)},0}=0$ is largest for $0.6 < z < 0.7$
while it significantly decreases for the higher redshifts. It gives us a
possibility to discriminate between the Perlmutter
model and brane models when the data from high-redshift
supernovae is available. On the other hand,
the difference between the best-fit Perlmutter model with
a $\Lambda$-term \cite{Perlmutter99} and the best-fit brane models becomes
detectable for redshifts $z > 1.2$. It is interesting
that brane models predict brighter galaxies for such redshifts which
is in agreement with the measurement of the $z = 1.7$ supernova \cite{Riess01}.
In other words, if the farthest $z >1$ supernovae were brighter, the brane
universes is allowed.
 
One should note that we made our analysis without excluding any supernovae from the sample.
However, from the formal point of view, when we analyze the full sample A,
all models should be rejected even on the confidence level of 0.99.
One of the reason could be the fact that the assumed error are
too small. However, in majority of papers another solution is suggested.
Usually, one excludes 2 supernovae as outliers, and 2 as reddened from the sample
of 42 high-redshift supernovae and eventually 2 outliers from the sample of 18
low-redshift supernovae.
We decided to use the full sample A as our basic sample
because a rejection of any supernovae from the sample can be the source
of lack of control for selection effects. However, for completness, we also
made our analysis using samples B and C.
It emerged that it does not significantly changes our results, though,
increases quality of the fit.
Formally, the best fit for the sample B (56 supernovae) is
($\chi^2=57.3$): $\Omega_{k,0}=-0.1$
$\Omega_{m,0}=0.17$, $\Omega_{\lambda,0}=0.06$, $\Omega_{\Lambda_{(4)},0}=0.87$.
For the flat model we obtain ($\chi^2=57.3$):
$\Omega_{m,0}=0.12$, $\Omega_{\lambda,0}=0.06$,
$\Omega_{\Lambda_{(4)},0}=0.82$,
while for a "realistic" model ($\Omega_{m,0}=0.25$, $\Omega_{\lambda,0}=0.02$)
$\chi^2=57.6$.
Formally, the best fit for the sample C (54 supernovae)
($\chi^2=53.5$) gives $\Omega_{k,0}=0$ (flat)
$\Omega_{m,0}=0.21$, $\Omega_{\lambda,0}=0.04$,
$\Omega_{\Lambda_{(4)},0}=0.75$,
while for "realistic" model ($\Omega_{m,0}=0.27$, $\Omega_{\lambda,0}=0.02$)
$\chi^2=53.6$.
 
One should note that we have also separately estimated the value of ${\cal M}$
for sample B and C.
We obtained ${\cal M}=-3.42$ which is again in a good agreement with
the results of \cite{Efstathiou99} (for a "combined" sample one
obtains ${\cal M}=-3.45$). However, if we use this value in our analysis
it does not change significantly the results ($\chi^2$ does not change more than
$1$ which is a marginal effect for $\chi^2$ distribution for 53 or 55 degrees
of freedom).
 
\begin{figure}[h]
\includegraphics[width=0.8\textwidth]{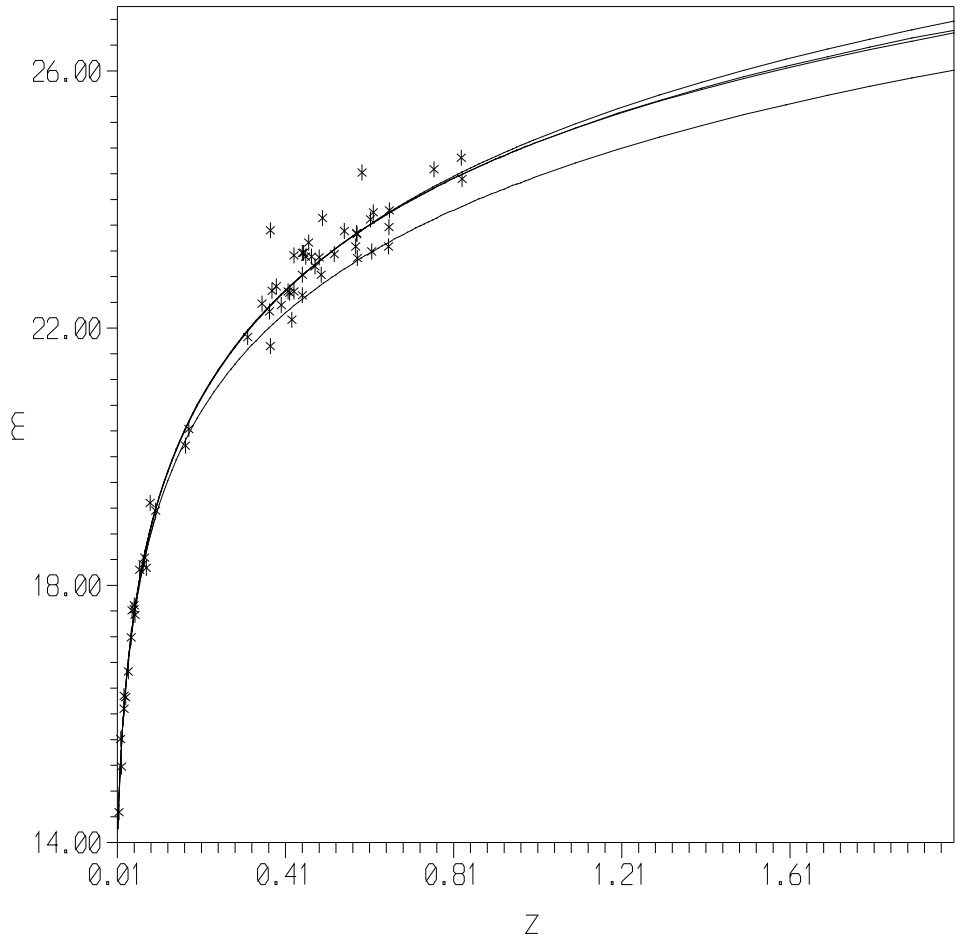}
%\includegraphics[scale=.31]{ff4.eps}
%\centerline{\epsfxsize=4cm\epsfbox{zdabnow.ps}}
\caption{The redshift-magnitude relation for $\gamma = -1/3$ brane universes
(phantom matter on the brane). The top line is the Perlmutter model and
the bottom line is the Einstein-de Sitter model. In the middle are two
overlapping line for the best-fitted and best-fitted flat brane models.}
\label{Fig.4}
\end{figure}
 
In Fig. \ref{Fig.4} we present a redshift-magnitude relation (\ref{chir1}) for
brane models with dark energy ($\gamma = -1/3$) and $\Omega_{\cal U}>0$.
To obtain an acceptable fit $\Omega_{{\cal U},0}$ should be so large
as $ \simeq 0.2$.
Note that the theoretical curves are very close to that of the Perlmutter
model \cite{Perlmutter99} which means that the dark energy
cancels the positive-pressure influence of the $\varrho^2$ term and can simulate
the negative-pressure influence of the cosmological constant to cause
cosmic acceleration. From the formal point of view the best fit (Tab.II) is
($\chi^2=95.4$) for $\Omega_{k,0}= 0.2$, $\Omega_{d,0}=0.7$, $\Omega_{\lambda,0}=-0.1$,
$\Omega_{{\cal U},0} = 0.2$, $\Omega_{\Lambda_{(4)},0}=0$ which
means that the cosmological constant must necessarily {\it vanish}. From
this result we can conclude that the dark energy $p = - (4/3) \varrho$ can {\it
mimic} the contribution from the $\Lambda_{(4)}$-term in standard models.
For the best-fit flat model ($\Omega_{k,0}=0$) we have
($\chi^2=95.4$): $\Omega_{d,0}=0.2$, $\Omega_{\lambda,0}=-0.1$,
$\Omega_{{\cal U},0} = 0.2$, $\Omega_{\Lambda_{(4)},0}=0.7$.
 
However if we excluded possibility that $\Omega_{\lambda,0}<0$, than
for value of the parameter $\Omega_{\lambda,0}=0.01$ we obtain:
$\Omega_{k,0}= 0.2$, $\Omega_{d,0}=0.5$, $\Omega_{\lambda,0}=0.01$,
$\Omega_{{\cal U},0} = 0.2$, $\Omega_{\Lambda_{(4)},0}=0,09$
For the best-fit flat model ($\Omega_{k,0}=0$) we have
($\chi^2=95.5$): $\Omega_{d,0}=0.05$, $\Omega_{\lambda,0}=0.01$,
$\Omega_{{\cal U},0} = 0.2$, $\Omega_{\Lambda_{(4)},0}=0,74$
which means that the cosmological constant is not {\it vanish} in such type of
model.
 
\begin{table}
\noindent
\caption{Results of the statistical analysis for the dark energy (phantom
matter) on the brane for Perlmutter Sample A with $\Omega_{{\cal U},0}=0.2$
Two upper line are best fit model and best fit flat model for sample.
The next two line are the model with $\Omega_{\lambda,0}=0.01$
}
\begin{tabular}{@{}p{1.5cm}p{0.5cm}rrrrr}
\hline
Sample & N & $\Omega_{k,0}$ & $\Omega_{\lambda,0}$ & $\Omega_{d,0}$ &
$\Omega_{\Lambda,0}$ & $\chi^2$ \\
\hline
  A   &  60 &  0.2 & -0.10 & 0.70 & 0.00 & 95.4\\
      &     &  0.0 & -0.10 & 0.20 & 0.70 & 95.4\\
      &     &  0.2 &  0.01 & 0.50 & 0.09 & 95.5\\
      &     &  0.0 &  0.01 & 0.05 & 0.74 & 96.5\\
\end{tabular}
\end{table}

\begin{figure}[h]
\includegraphics[angle=0,width=0.8\textwidth]{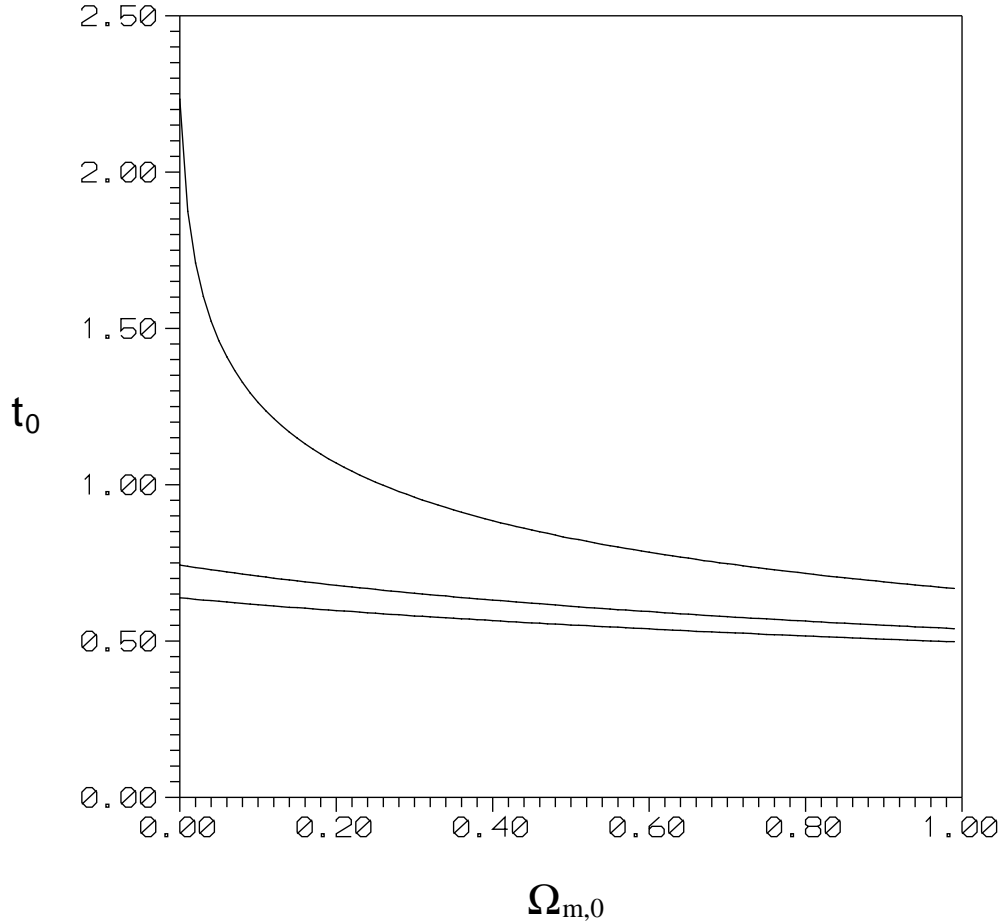}
%\includegraphics[angle=270,scale=.31]{ff5.eps}
%\centerline{\epsfxsize=4cm\epsfbox{prl2.ps}}
\caption{The age of the universe $t_0$ in units of $H_0^{-1}$
for the brane models with dust ($0 \leq \Omega_{m,0} \leq 1$
on the horizontal axis). Here $\Omega_{{\cal U},0} = \Omega_{k,0} = 0$,
$\Omega_{{\lambda},0} = 0, 0.05, 0.1$ (top, middle, bottom).}
\label{Fig.5}
\end{figure}
 
\section{Other tests for brane cosmology}
 
\subsection{Brane models and age of the universe}
 
\begin{figure}[h]
\includegraphics[angle=0,width=0.8\textwidth]{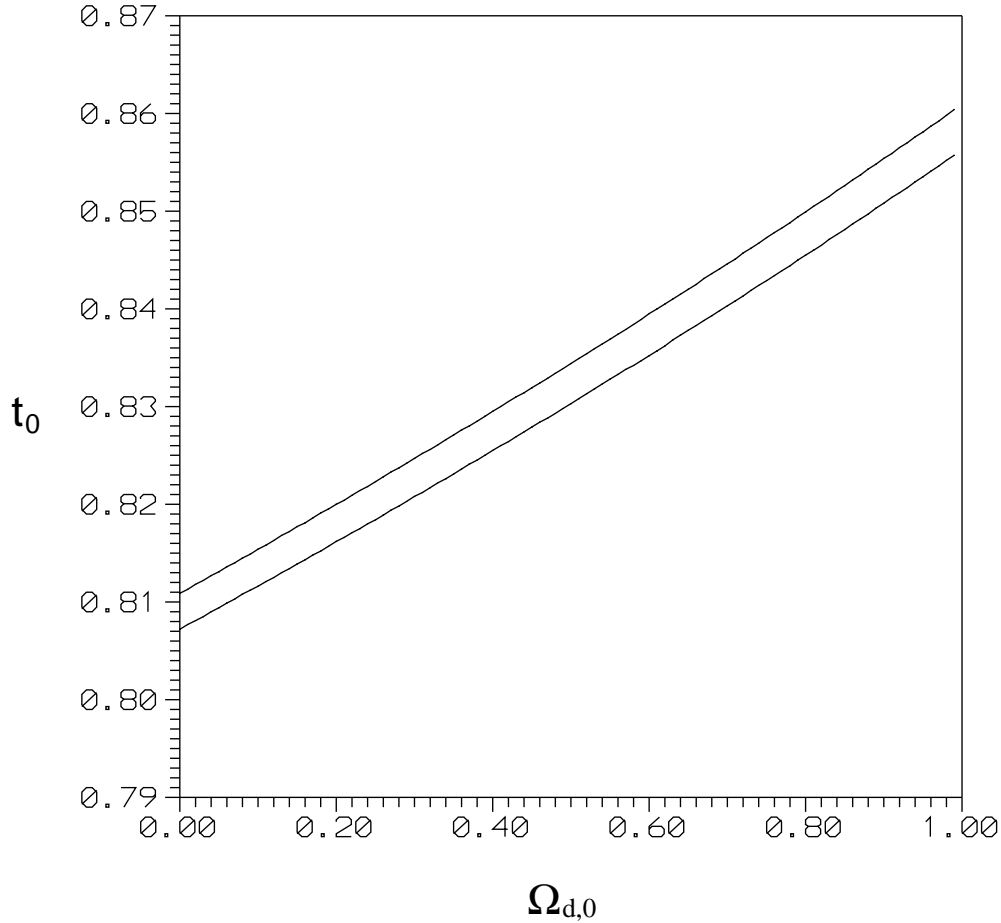}
%\includegraphics[angle=270,scale=.31]{ff6.eps}
%\centerline{\epsfxsize=4cm\epsfbox{prl6.ps}}
\caption{The age of the universe $t_0$ in units of $H_0^{-1}$ for the
brane models with dark energy (phantom matter) on the brane
($0 \leq \Omega_{d,0} \leq 1$ on the horizontal axis). Here $\Omega_{{\cal U},0} =
0.2$, $\Omega_{{\lambda},0} = 0.05, 0$ (top, bottom) which shows weaker influence
of the brane effects to increase the age.}
\label{Fig.6}
\end{figure}

Now let us briefly discuss the effect of brane parameters and dark energy onto
the age of the universe which according to (\ref{eq:FriedCCC}) is given by
\bea
\label{age}
H_0 t_0 = \int_0^1 \left\{ \Omega_{\gamma,0} x^{-3\gamma +
4} + \Omega_{{\lambda},0} x^{-6\gamma + 4}
\right. \nonumber \\ \left.
+ \Omega_{{\cal U},0} +
\Omega_{k,0} x^2 + \Omega_{\Lambda_{(4)},0} x^4 \right\}^{-\frac{1}{2}} x dx  ,
\eea
where $x = a/a_0$. We made a plot for the dust $\gamma = 1$ on the brane in
Fig. \ref{Fig.5} which shows that the effect of quadratic in
energy density term represented by $\Omega_{\lambda}$ is to {\it
lower} significantly the age of the universe. The problem can be
avoided, if we accept the dark energy $\gamma = - 1/3$
on the brane \cite{darkenergy} , since the dark energy
has a very strong influence to increase the age.
In Fig. \ref{Fig.6} we made a plot for this case which shows how
the dark energy enlarges the age.

\subsection{Angular diameter versus redshift for brane models}
 
Finally, let us study the angular diameter test for brane universes.
The angular diameter of a galaxy is defined by
\be
\label{angdiam}
\theta = \frac{d(z+1)^2}{d_{L}} ,
\ee
where $d$ is a linear size of the galaxy. In a flat dust ($\gamma =1$)
universe the angular diameter $\theta$ has the minimum value $z_{min} = 5/4$.
It is particularly interesting to notice that for flat
 brane models with $\Omega_{\lambda,0} \approx 0, \Omega_{\Lambda_{(4)}}
\approx 0$ the dark radiation shift the minimum of $\theta(z)$
relation towards to higher $z$ for $\Omega_{{\cal U},0} \leq 0$
while the ordinary radiation shift this minimum towards to lower $z$.
It should be also noted that dark radiation decrease the value
of $\theta(z_{min})$ for $\Omega_{{\cal U},0} \leq 0$
while the ordinary radiation increase this value.
This is a general influence of negative dark radiation
onto the angular diameter size for brane models.

\begin{figure}[h]
\includegraphics[width=0.8\textwidth]{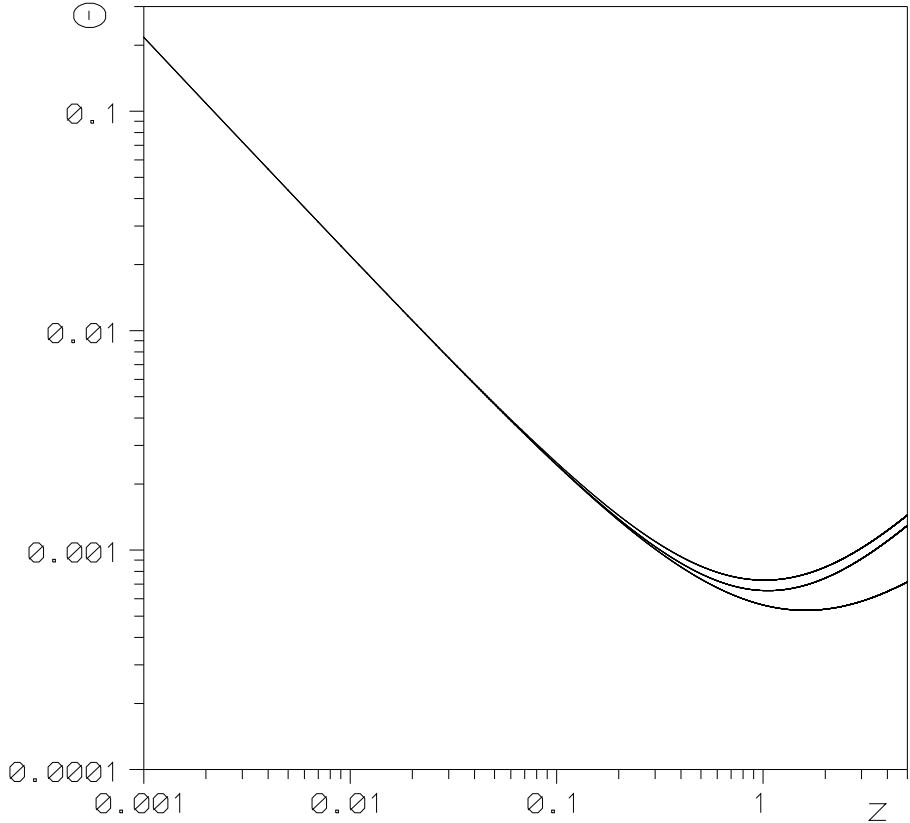}
%\includegraphics[scale=.31]{ff7.eps}
%\centerline{\epsfxsize=4cm\epsfbox{figopis5.ps}}
\caption{The angular diameter $\theta(z)$ for
$\Omega_{\lambda,0} = 0.1, \Omega_{m,0} = 0.3, \Omega_{\Lambda_{(4)}} = 0.72$,
and the two values of $\Omega_{{\cal U},0} = 0.1, -0.1$ (top, middle)
in comparison with the model of Perlmutter with $\Omega_{m,0} = 0.28,
\Omega_{\Lambda_{(4)}} = 0.72$ (bottom).}
\label{Fig.7}
\end{figure}

More detailed analytic and numerical studies $\theta(z)$ relation \cite{Dabreg02}
show that the increase of $z_{min}$ is even more sensitive to negative
values of $\Omega_{\lambda,0}$ ($\varrho^2$ contribution). Similarly
as for the dark radiation $\Omega_{{\cal U},0}$, the minimum disappears
for some large negative $\Omega_{{\lambda},0}$.
Positive $\Omega_{{\cal U},0}$ and $\Omega_{{\lambda},0}$ make $z_{min}$ decrease.
In Fig. \ref{Fig.7} we present a plot from which one can see the
sensitivity of $z_{min}$ to $\Omega_{{\cal U},0}$. We have also checked
\cite{Dabreg02} that the dark energy $\Omega_{d,0}$ (phantom matter) has very
little influence onto the value of $z_{min}$.
 
\section{Conclusions}
 
We shown that there exists an effective method of constraining exotic physics
coming from superstrings M theory from observation of distant supernovae.
We obtain the estimated value the density parameters $\Omega_{\lambda,0}$
and $\Omega_{\Lambda,0}$ .
 
Finally, as a result we also obtain that at high redshifts the
expected luminosity of supernovae Ia should be brighter then in the Perlmutter
model. For the best fit value we obtain
$\Omega_{\lambda,0} \simeq 0.01$ which seems to be unrealistic.
It is because if we consider pure Randall-Sundrum
models, then there is a constraint on the parameter $\Omega_{\lambda,0}$
from the requirement of not violating the four-dimensional gravity on sufficiently
large spatial scale. This constraint is that
value of $\lambda$ is to be no less than about $(100 \, GeV)^4$, which means that
during the late epoch the $\rho^2$ term in the model should be small.

So, the obtained value of $\Omega_{\lambda,0} \sim 0.01$ is the
observational limit which is not based on theoretical model assumptions.
The density $\rho_{m,0}$ at the time relevant for supernovae
measurements is about $(10^{-3}eV)^4$. Thus, the size of the parameter
$\Omega_{\lambda,0}$ is on purely theoretical grounds, at most $10^{-56}$.
The fits discussed in the paper innvolve $\Omega_{\lambda,0}$ of order $0.01$.
Therefore similarly to the cosmological constant problem there is a coincidence
problem with brane tension $\lambda$ (it is treated as a constant) namely:
why don't we see the large brane tension expected from the Randall-Sundrum theory
which is about $10^{54}$ times larger than the value predicted by the Friedmann
equation which fit SNIa data. A phenomenological solution to this problem seems
to be dynamically decaying $\lambda$.
 
\section{Acknowledgements}
 
The author are very grateful Prof M.D\c{a}browski for his comments and
interesting remarks.

\end{document}